\def\papertitle{ICGAN: An implicit conditioning method for interpretable feature control of neural audio synthesis}
\def\paperauthorA{Yunyi Liu}
\def\paperauthorB{Craig Jin}

\documentclass[twoside,a4paper]{article}
\usepackage{etoolbox}
\usepackage{dafx_24}
\usepackage{amsmath,amssymb,amsfonts,amsthm}
\usepackage{euscript}
\usepackage{subcaption}
\usepackage[T1]{fontenc}
\usepackage[utf8]{inputenc}
\usepackage{multirow}
\usepackage{multicol}
\usepackage{booktabs}
\usepackage{tikz}
\usepackage{ifpdf}
\usepackage[english]{babel}
\usepackage{caption}
\usepackage{color}

\input glyphtounicode
\ninept

\newcounter{numauth}
\newcounter{listcnt}
\newcommand\authcnt[1]{\ifdefined#1 \stepcounter{numauth} \fi}

\newcommand\addauth[1]{
\ifdefined#1 
\stepcounter{listcnt}
\ifnum \value{listcnt}<\value{numauth}
\appto\authorslist{, #1}
\else
\appto\authorslist{~and~#1}
\fi
\fi}

\authcnt{\paperauthorB}

\def\authorslist{\paperauthorA}
\addauth{\paperauthorB}
\usepackage{times}

\newif\ifpdf
\ifx\pdfoutput\relax
\else
   \ifcase\pdfoutput
      \pdffalse
   \else
      \pdftrue
\fi

\usepackage[pdftex,
    pdftitle={\papertitle},
    pdfauthor={\authorslist},
    pdfsubject={Proceedings of the 27th International Conference on Digital Audio Effects (DAFx24)},
    colorlinks=false, % links are activated as color boxes instead of color text
    bookmarksnumbered, % use section numbers with bookmarks
    pdfstartview=XYZ % start with zoom=100% instead of full screen; especially useful if working with a big screen :-)
]{hyperref}
\title{\papertitle}

\twoaffiliations{
\paperauthorA \,\thanks{\vspace{-3mm}}}
{\href{https://dafx24.surrey.ac.uk}{Computing and Audio Research Lab} \\ University of Sydney\\ Sydney, Australia\\
{\tt \href{mailto:yunyi.liu@sydney.edu.au}{yunyi.liu@sydney.edu.au}}
}
{\paperauthorB \,}
{\href{https://dafx23.create.aau.dk/}{Computing and Audio Research Lab} \\ University of Sydney \\ Sydney, Australia \\ {\tt \href{mailto:craig.jin@sydney.edu.au}{craig.jin@sydney.edu.au}}
}

\begin{document}
% more pdf-tex settings:
\ifpdf % used graphic file format for pdflatex
  \DeclareGraphicsExtensions{.png,.jpg,.pdf}
\else  % used graphic file format for latex
  \DeclareGraphicsExtensions{.eps}
\fi

%\makeatletter
%\pdfbookmark[0]{\@pdftitle}{title}
%\makeatother

\maketitle

\begin{abstract}
Neural audio synthesis methods can achieve high-fidelity and realistic sound generation by utilizing deep generative models. Such models typically rely on external labels which are often discrete as conditioning information to achieve guided sound generation. However, it remains difficult to control the subtle changes in sounds without appropriate and descriptive labels, especially given a limited dataset. This paper proposes an implicit conditioning method for neural audio synthesis using generative adversarial networks that allows for interpretable control of the acoustic features of synthesized sounds.  Our technique creates a continuous conditioning space that enables timbre manipulation without relying on explicit labels. We further introduce an evaluation metric to explore controllability and demonstrate that our approach is effective in enabling a degree of controlled variation of different synthesized sound effects for in-domain and cross-domain sounds.
\end{abstract}

\section{Introduction}
\label{sec:intro}

In recent years, neural audio synthesis has achieved excellent performance in speech modelling~\cite {wavenet, MelGAN} and music generation~\cite{RAVE, GANsynth}. However, it remains difficult to model sound effects with pure generative models in that neural networks typically offer a low degree of interpretability and controllability. A common approach to achieving sound generation control is via conditioning on external information. Conditional generation is a common technique to constrain generative models to generate data based on specific input information, which can be categorical labels, texts, images, compressed embeddings, etc. Considering the limited availability of datasets for general sound effects, conditioning on discrete labels can be effective, as most datasets provide at least categorical information. For example, Barahona-Ríos and Collins~\cite{conwavegan} proposed to synthesize knocking sounds conditioned on different emotions to guide the GAN generator for audio generation. Similarly, Communita et al.~\cite{neuralfootstep} also studied GAN sound generation for modelling footstep sounds conditioned on discrete contact surfaces while Liu et al~\cite{liu2023conditional} focused on conditioning on different types of audio. To be more effective in utilizing available audio datasets, DarkGAN~\cite{DarkGAN} uses knowledge distillation (KD) to extract categorical information from large pre-trained classification models~\cite{PANNs}. The classification model is able to output the associated probabilities (termed as soft labels) for each class, which could be utilized as conditioning information to guide the sound synthesis. Although such approaches achieved higher audio qualities without requiring large amounts of labelled datasets, the discrete nature of the conditioning method still poses drawbacks, including limited expressiveness~\cite{10.5555/3157382.3157633, constraints}, lack of continuity~\cite{Vrtes2018FlexibleAA}, and failure to capture the hierarchical semantic relationships between classes~\cite{hierarchical}. Conditioning on text embeddings, on the other hand, provides a simple yet effective control with human languages. AudioLDM~\cite{AudioLDM} for example, is capable of synthesizing intricate audio waveforms agnostic to sound types given input texts by training in a self-supervised fashion which requires less label data. Nevertheless, text-to-audio synthesis models still require large amounts of audio datasets with descriptive labels in order to achieve reasonable performance. Additionally, texts themselves are also limited by the expressiveness in describing the subtle differences among different sounds. 
\\
\\
Apart from conditioning on discrete labels, it is possible to use continuous vectors~\cite{conaug} as conditioning information, but may require specific training setup. Regression labels (continuous, such as ages and angles in image) provide a smooth and expressive representation that can help capture hierarchical~\cite{CCGAN} and nuanced variations across different classes~\cite{PcDGAN}. CcGAN~\cite{CCGAN} achieves this by assigning weights using a Gaussian kernel based on the distance from the target label, allowing the model to handle sparse data more effectively. The proposed hard vicinal and soft vicinal discriminator loss allows the generative model to smooth the transition across continuous labels. It has been demonstrated that by introducing uncertainty, label smoothing~\cite{Labelsmoothing} helps generative models improve their training stability and robustness. However, in the domain of audio, it is generally very difficult to obtain regression labels, and the number of high-quality datasets for sound effects pales in comparison with images. This makes applying a similar conditioning method to sound effects modelling difficult. 
\\
\\
In this research, we study sound effects modelling under a GAN architecture by conditioning our generator on probabilistic soft labels. Different from CcGAN~\cite{CCGAN}, we wish to blur the boundaries between discrete classes without requiring regression labels. To achieve this, we introduce an implicit conditioning method by manipulating discrete labels into continuous probabilistic vectors as conditioning. In Section~\ref{Method}, we elaborate on the proposed method by integrating it into a WGAN~\cite{WGAN} model for Mel-spectrogram generation. In Section\ref{sec: experiments} and Section\ref{sec: evaluation}, we discuss the training details and the experiments as well as the evaluation metrics we used. We show the control effectiveness of the soft-labelling approach and propose two metrics to understand and evaluate the controllability of our conditioning approach using a pre-trained audio classifier. The results are shown in Section~\ref{sec: Results}.

\begin{figure*}[h]
  \centering
  \centerline{\includegraphics[width=\textwidth]{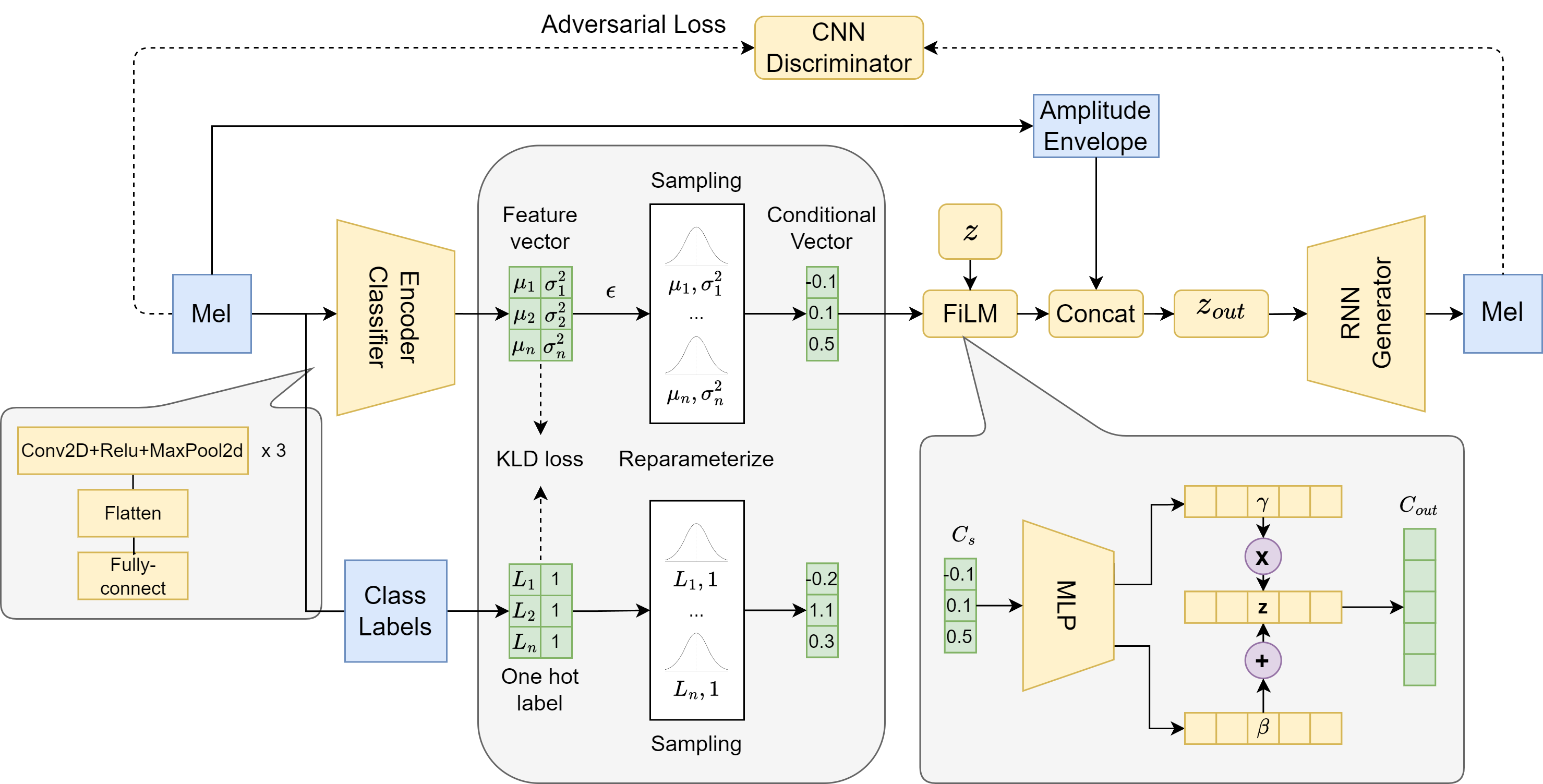}}
  \caption{Proposed model architecture. The model is composed of a CNN encoder classifier, an RNN generator, and a CNN discriminator. Yellow boxes are part of the neural networks. Blue boxes indicate training inputs and outputs, while greens represent explicit variables.}
%  \vspace{1.5cm}
  \label{fig: model}
\end{figure*}

\section{Methodology}
\label{Method}

% \begin{figure}[ht]
%   \centering
%   \centerline{\includegraphics[width=\linewidth]{vanilla.png}}
%   \caption{A conventional class-conditional GAN}
% %  \vspace{1.5cm}
%   \label{fig:conditional}
% \end{figure}

% There are several ways to incorporate external information as conditioning information, such as via channel concatenation~\cite{congan}, element-wise multiplication, and affine transformations~\cite{FILM}. In Figure~\ref{fig:conditional}, we show a classical conditioning method by vector concatenation in WaveGAN~\cite{neuralfootstep} for sound effects modelling. The sounds are separated into $n_c$ number of classes with a one-hot vector $L$ with a dimension of $n_c$ which encodes the categorical information. The embedding representations of $L$ are then concatenated with the latent representations $z$. In this way, the one-hot vectors are utilized explicitly as conditioning information for both the generator and discriminator components, which should facilitate a more structured and informative learning environment. However, because of the limited expressivity and difficulty in capturing semantic information between discrete classes as previously discussed in Section~\ref{sec:intro}, we propose to condition the generator on probability scores output from a separate classifier as illustrated in Figure~\ref{fig: model}.

\subsection{Proposed conditioning method}

We start by formulating our research goal. For a generator model, $G$, whose task is to output a two-dimensional Mel-spectrogram, $M$, we wish to condition the model on a continuous vector, $C$, which encodes the acoustic attributes of input sounds across different categories. To this end, we replace the one-hot vectors with a random variable sampled from a Gaussian distribution. Given a conditional vector \(C_s\), sampled from a Gaussian distribution parameterized by mean and variance, our modelling objective is to estimate the conditional probability distribution \(p(x | C_s)\) of the target data \(x\). This can be formulated as:

\begin{equation}
    p(x | C_s) = \int p(x | z, C_s) \, p(z | C_s) \, dz
\end{equation}

where:
\begin{itemize}
    \item \(x\) represents the target data we aim to model,
    \item \(C_s\) is the conditioning vector, providing context or attributes that the target data \(x\) should adhere to, and \\
    $C_s \sim \mathcal{N}(\mu, \sigma^2)$, where $\mu$ and $\sigma$ are learned from a classifier.
    \item \(z\) is a latent variable capturing aspects of \(x\) not specified by \(C_s\),
    \item \(p(z | C_s)\) represents the distribution of latent variables conditioned on \(C_s\),
    \item \(p(x | z, C_s)\) denotes the likelihood of \(x\) given both \(z\) and \(C_s\).
\end{itemize}

Specifically, we first employ a CNN-based encoder classifier as shown in Figure~\ref{fig: model} to extract acoustical information from $M$. The encoder classifier learns the spectral-temporal information and is expected to output class probabilities of the inputs. Instead of using the probabilities directly as conditioning labels, we wish to obtain a control space that allows us to interpolate the sound characteristics smoothly. To this end, we transform the output logits from the encoder classifier into mean and variance variables, $\mu$ and $\sigma^2$, respectively, which are used for re-parameterization to enable gradient flow. Similar to a VAE\cite{VAE}, we sample a vector from standard Gaussian distribution as $\mathbf{\epsilon} \sim \mathcal{N}(0, \mathbf{I})$ to obtain the sampled vector $C_s = \mu + \mathbf{\epsilon} \cdot \sigma^2$. By introducing uncertainty and noise through the sampling process, we wish to construct a more continuous space of categorical information among different classes of sounds. Instead of relying on a simple concatenation process, we integrate the conditioning information via a feature-wise linear modulation (FiLM)~\cite{FILM} operation that is applied to both $C_s$ and $z$. FiLM applies an affine transformation to the intermediate features of a neural network using scaling and shifting operations, mathematically represented as $FiLM(z, C_s) = f_{\gamma}(C_s) \cdot z + f_{\beta}(C_s)$, where $f_{\gamma}$ and $f_{\beta}$ are neural networks (typically MLPs) that map the conditioning input $C_s$ to the latent space dimension. This technique allows for precise control over how external information influences network activation values, leading to accurate and context-aware outputs. 

% In this way, $z$ is conditioned on a sampled vector with its mean $\mu$ and variance $\sigma^2$ learned from the classifier. 

\subsection{Model architecture}

To demonstrate how our conditioning method performs, we integrate our conditioning method into a GAN network for Mel-spectrogram generation. Mel-spectrogram correlates well with human perception and has received much attention in neural audio synthesis~\cite{VQVAE, Specsingan}. It serves as an interpretable intermediate representation that could be effectively learned with small amounts of data. With the advancement of neural vocoders~\cite{MelGAN, Hifigan}, we can usually get decent audio reconstructions from Mel-spectrograms. We denote our model as ICGAN (implicit conditioning GAN) with the entire model architecture as shown in Figure~\ref{fig: model}. In addition to conditioning the generator on the class labels, we also condition it on the extracted amplitude envelope from the Mel-spectrograms. Amplitude envelopes serve as a regulator that guides the generator to synthesize spectrograms with similar envelopes. This enables us to synthesize audio in a frame-level manner by inputting a guiding envelope during the inference. Additionally, with the help of amplitude information, the generator model may achieve higher synthesis quality, which is discussed in Section~\ref{sec: Results}. To obtain the amplitude envelope from the Mel-spectrograms, we sum and average the power amplitude across all frequency bins to get the frame-level amplitude envelope $A$:
\begin{equation}
\label{eq: envelope}
    A(t) = \frac{1}{F} \sum_{f=1}^{F} P(f, t) \,\, \text{,}
\end{equation}
where $P(f, t)$ is the power at frequency bin $f$ and time frame $t$ for all frequency bins $f \in [1,F]$. In this way, the generator is guided by the amplitude information as well as the class-conditioning vectors. To synthesize the 2D spectrograms, we use the same RNN-based generator to model Mel-spectrograms as ~\cite{STGAN}. The generator is a three-layer GRU unit with an internal size of 512 followed by a linear layer. It sequentially transforms the input information to Mel-spectrograms with the same sequence length. As for the encoder classifier and the discriminator, we adapt a simple LeNet-5~\cite{LENET} as shown in Figure~\ref{fig: model}. Both the encoder classifier and discriminator models employ 2D CNN layers to extract spatial information from the input spectrograms. Once the model is trained, we transform the generated Mel spectrogram to audio waveforms using a pre-trained neural vocoder~\cite{Hifigan}. The HifiGAN vocoder was trained on the Audioset\cite{Audioset} dataset, which contains millions of labelled sound events. It is expected to be able to convert Mel-spectrograms back to audio waveforms of agnostic to the audio source. Our complete model implementation can be accessed at the companion website\footnote{\url{https://github.com/Reinliu/ICGAN}}.

\subsection{Loss functions}

Instead of providing explicit labels as input into the generator, we employ continuous categorical labels as conditioning vectors learned from the classifier. To facilitate the learning of the conditioning vectors, we force the re-parameterized Gaussian distribution to be similar to a Gaussian distribution formulated based on discrete class labels with a ground-truth mean and variance pair (refer to Fig.~\ref{fig: model}). We use the Kullback-Leibler divergence (KLD) to measure the similarity of the two distributions. We compute a regularization loss $L_{reg}$ that incorporates the KLD using the one-hot vector label $L$, as well as the mean $\mu_n$ and variance $\sigma_n^{2}$ vectors obtained from the encoder classifier:

\begin{equation}
\label{eq: KLD_loss}
    \mathcal{L}_{reg} = -\frac{1}{2} \sum_{n=1}^{N} \left(1 + \log(\sigma_{n}^{2}) - (\mu_{n}-L_n)^{2} - \sigma_{n}^{2}\right),
\end{equation}
where $N$ is the dimension of the label space, which corresponds to the total number of classes, and one hot values $L_n$ are used to indicate the target distribution mean. Apart from the regularization loss, we adopt the Wasserstein GAN (WGAN)~\cite{WGAN} loss framework for the discriminator, mirroring its proven approach to evaluate the discrepancy between real and generated data distributions. The generator loss is shown below:

\begin{equation}
L_G = -\mathbb{E}_{\tilde{x} \sim \mathbb{P}_g} [D(\tilde{x})] + \mathcal{L}_{reg}
\end{equation}
where $\mathbb{E}_{\tilde{x} \sim \mathbb{P}_g} [\cdot]$ denotes the expectation over samples $\tilde{x}$ generated by the generator's distribution $\mathbb{P}_g$ and $D(\tilde{x})$ represents the discriminator's (critic's) score for the generated samples. Notice that our generator loss also includes the regularization loss as illustrated above. In this way, the generator learns to accommodate changes associated with the classifier updating its output. This allows the generator and encoder to update their weights simultaneously within each batch, helping to stabilize the training and make the results more reliable.\\

With regard to the discriminator, we enhance training stability and enforce the Lipschitz constraint essential for optimal discriminator behavior by integrating the gradient penalty mechanism that is a hallmark of the WGAN-GP model~\cite{WGAN-GP} as shown below:

\begin{align}
    L_D = & \; \mathbb{E}_{\tilde{x} \sim \mathbb{P}_g} [D(\tilde{x})] - \mathbb{E}_{x \sim \mathbb{P}_r} [D(x)] \nonumber \\
    & + \lambda \cdot \mathbb{E}_{\hat{x} \sim \mathbb{P}_{\hat{x}}} \left[\left(\|\nabla_{\hat{x}} D(\hat{x})\|_2 - 1\right)^2\right]
\end{align}
\begin{itemize}
    \item $\mathbb{E}_{x \sim \mathbb{P}_r} [D(x)]$ is the average score that the discriminator gives to real samples.
    \item $\lambda \cdot \mathbb{E}_{\hat{x} \sim \mathbb{P}_{\hat{x}}} \left[\left(\|\nabla_{\hat{x}} D(\hat{x})\|_2 - 1\right)^2\right]$ represents the gradient \\
    penalty. This term enforces a Lipschitz constraint and is crucial for the model's stability. The penalty is applied to the gradients of the discriminator's scores with respect to interpolated samples between real and generated samples, pushing these gradients to have a norm of 1. We set $\lambda=10$ in our experiments.
\end{itemize}

\section{Experiments}
\label{sec: experiments}

\begin{figure*}[ht]
  \centering
  \begin{subfigure}[t]{0.3\linewidth}
    \caption{Discriminator loss}
    \includegraphics[width=\linewidth]{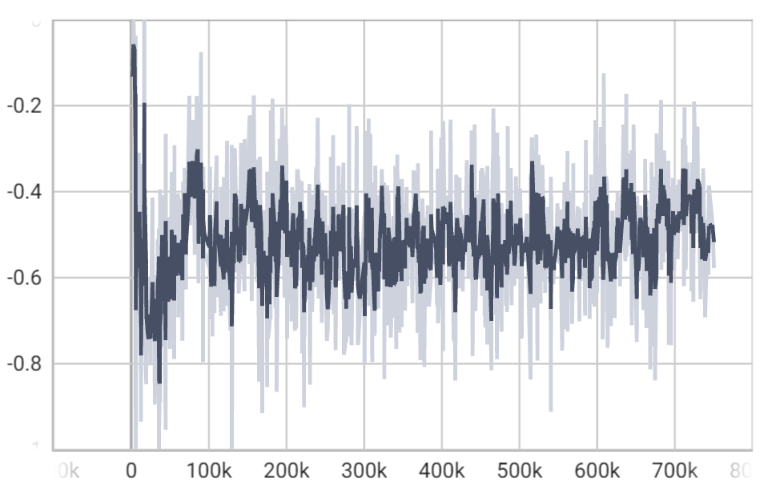}
    \label{fig: Disc loss}
  \end{subfigure}
  \hspace{0.5cm} % Space between the images
  \begin{subfigure}[t]{0.3\linewidth}
    \caption{Generator loss}
    \includegraphics[width=\linewidth]{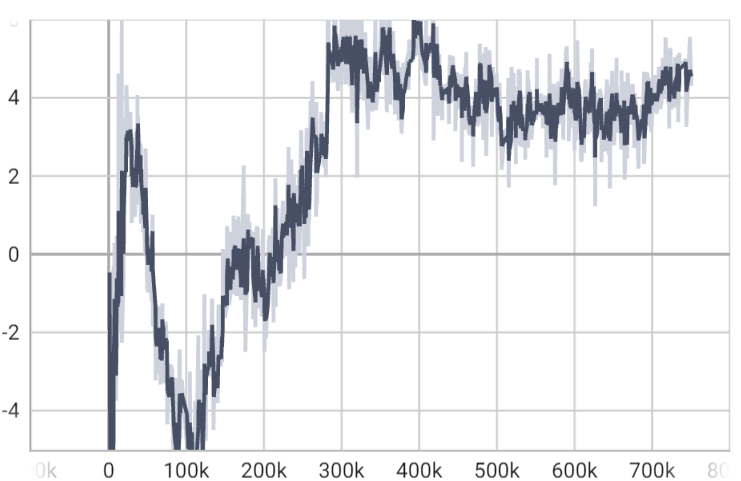}
    \label{fig: Gen loss}
  \end{subfigure}
  \hspace{0.5cm} % Space between the images
  \begin{subfigure}[t]{0.3\linewidth}
    \caption{Regularizer loss}
    \includegraphics[width=\linewidth]{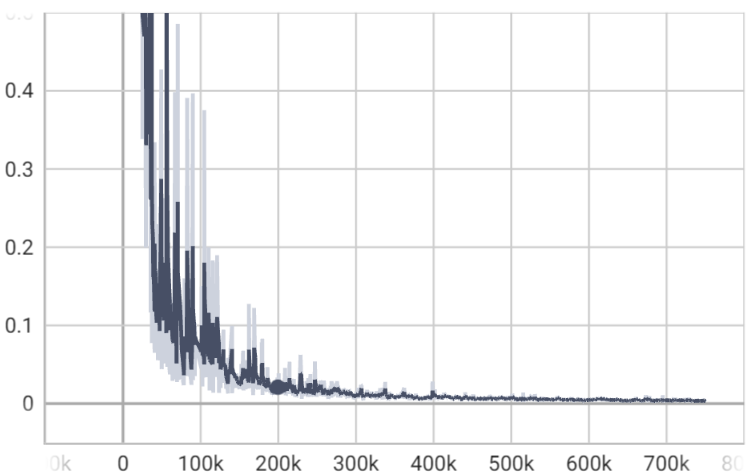}
    \label{fig: KLD loss}
  \end{subfigure}
  
  \caption{Visualization of the loss trend in our ICGAN model. We generally see a convergence after 300k iterations when all three losses were shown stable. The regularizer loss quickly converges while discriminator and generator loss oscillates around -0.5 and 4 respectively.}
  \label{fig: convergence}
\end{figure*}

\subsection{Dataset}
\label{sec: dataset}
As we are mainly interested in modelling sound effects, we meticulously construct an impact-sound-based audio dataset with sounds selected from the Boom Library\footnote{\url{https://www.boomlibrary.com/}} and BBC Sound Library~\cite{BBC}. Because of copyright issues, these datasets are not open to the public. Our dataset consists of three kinds of impact sounds: footsteps, gunshots, and hits. Each of these sound types are also comprised of several classes dependent on the labelling of such sounds. For example, footsteps contain 'Concrete', 'Gravel', 'Leaves', etc., while hits contain 'Punch face', 'Hit bag', 'Hit bone', etc. There are 6 subcategories of footstep sounds, 10 subcategories of gunshot sounds, and 9 categories of hit sounds. Each of the subcategories contains 100-200 sounds. In total, there are 4817 sounds and we split our dataset into 4335 (90\%) for training and 482 (10\%) for testing. Each of the sounds is sampled at 16~kHz and has a length of 4 seconds. We pre-process each sound to extract its Mel-spectrogram with a hop length of 64, and FFT size of 1024. All of the Mel-spectrograms have a length of 400 frames and 64 frequency bins. This setting corresponds to a pre-trained HiFiGAN vocoder~\cite{AudioLDM} trained on the Audioset~\cite{Audioset} dataset, allowing us to easily transform the Mel-spectrograms back to audio waveforms.

\subsection{Training}

We trained our model with the aforementioned datasets of three categories of sounds (footstep, gunshot, hits) for 10,000 epochs on an RTX 3080. Each model is conditioned with the variations within each category (eg. concrete, gravel, snow for the footstep category). We use an Adam optimizer~\cite{ADAM} with a learning rate of $l = 0.0001$, $\beta_1 = 0.5$, $\beta_2 = 0.999$. During training, the discriminator loss stabilizes between -0.4 and -0.7 after about 100~k iterations indicating convergence. The regularization loss converges more slowly and requires approximately 300~k iterations as shown in Figure~\ref{fig: convergence}. As the regularization loss is added to the generator loss, the generator loss oscillates until the regularization loss converges after the 300~k iterations. Eventually, after the regularization loss stabilizes, the generator loss converges to a value of around four. 
% loss demonstrates convergence, as evidenced by the reduction in variance over time and the establishment of a stable loss value in the later training phase.
\\
\\
In addition to the in-class sounds (sounds within the same category), we briefly test the inference capability of our model on cross-class sounds. Therefore, we further trained our model on a sound effect dataset DCASE2023~\cite{DCASE} and condition it on the 7 categories such as motor engine, dog barks, and rain. We show the performance of the conditioning method in Section~\ref{sec: Results}.

\section{Evaluation Methods}
\label{sec: evaluation}

\subsection{Interpolation between two classes}

% Main figure
\begin{figure*}[ht]
  \centering
  
  % Subfigure for footsteps
  \begin{subfigure}[t]{\linewidth}
    \centering
    % \begin{subfigure}[t]{0.18\linewidth}
    %   \includegraphics[width=\linewidth]{interpolation/footstep/footstep-class-0-4-6,9.png}
    % \end{subfigure}
    % \hspace{0.2cm} % Space between the images
    \begin{subfigure}[t]{0.22\linewidth}
      \includegraphics[width=\linewidth]{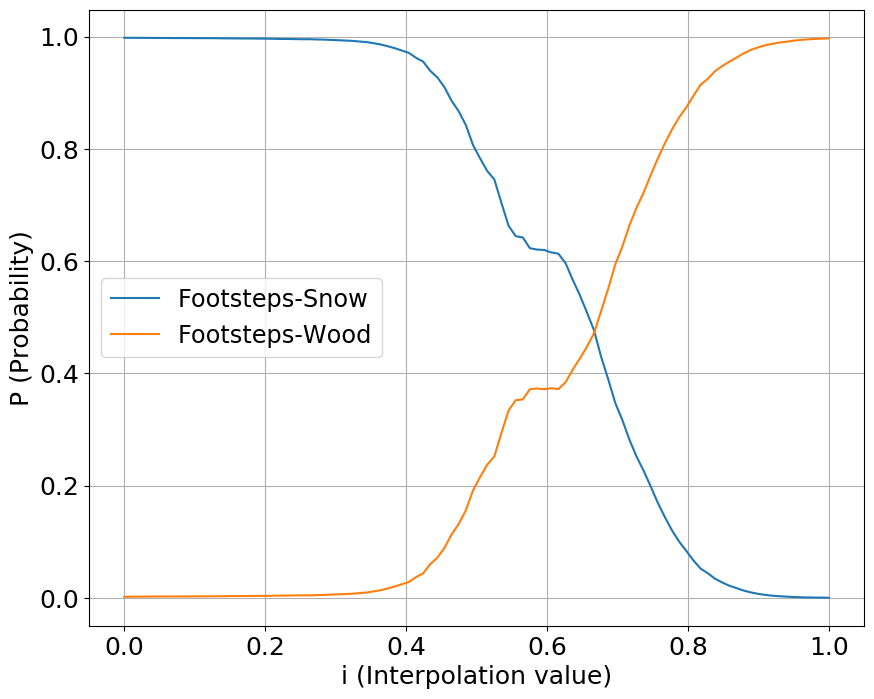}
    \end{subfigure}
    \hspace{0.2cm} % Space between the images
    \begin{subfigure}[t]{0.22\linewidth}
      \includegraphics[width=\linewidth]{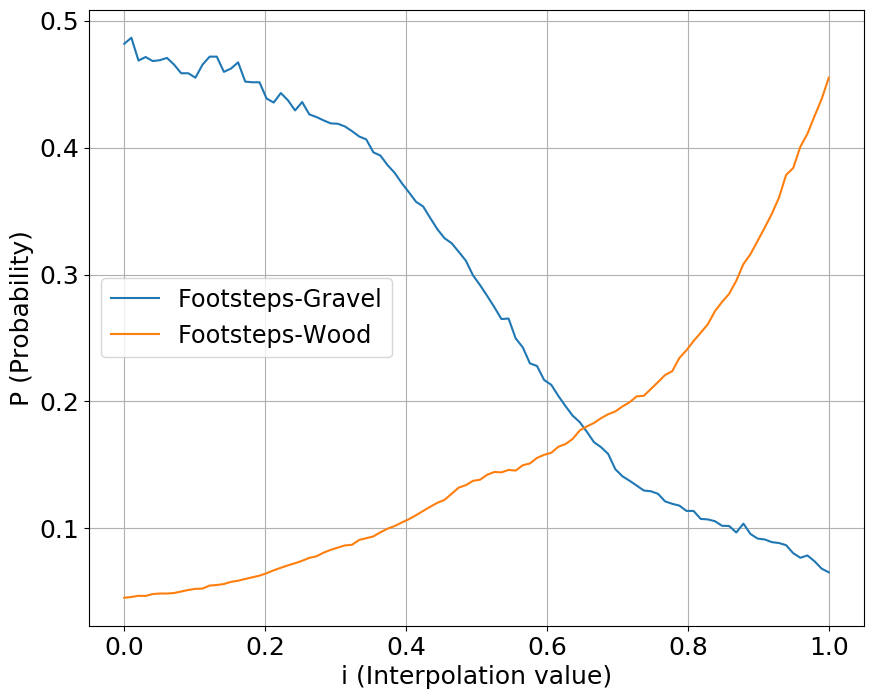}
    \end{subfigure}
    \hspace{0.2cm} % Space between the images
    \begin{subfigure}[t]{0.22\linewidth}
      \includegraphics[width=\linewidth]{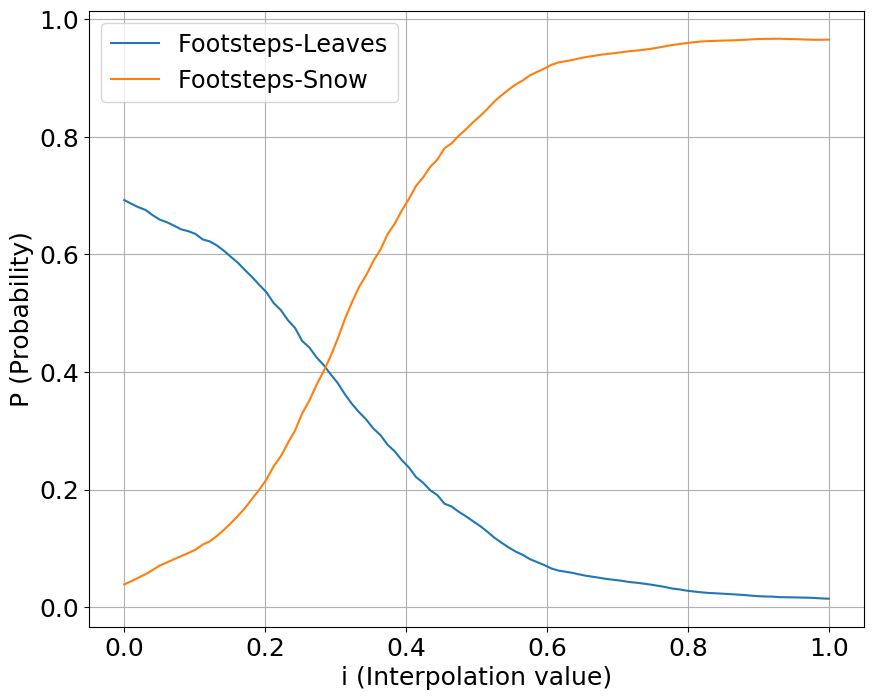}
    \end{subfigure}
    \hspace{0.2cm} % Space between the images
    \begin{subfigure}[t]{0.22\linewidth}
      \includegraphics[width=\linewidth]{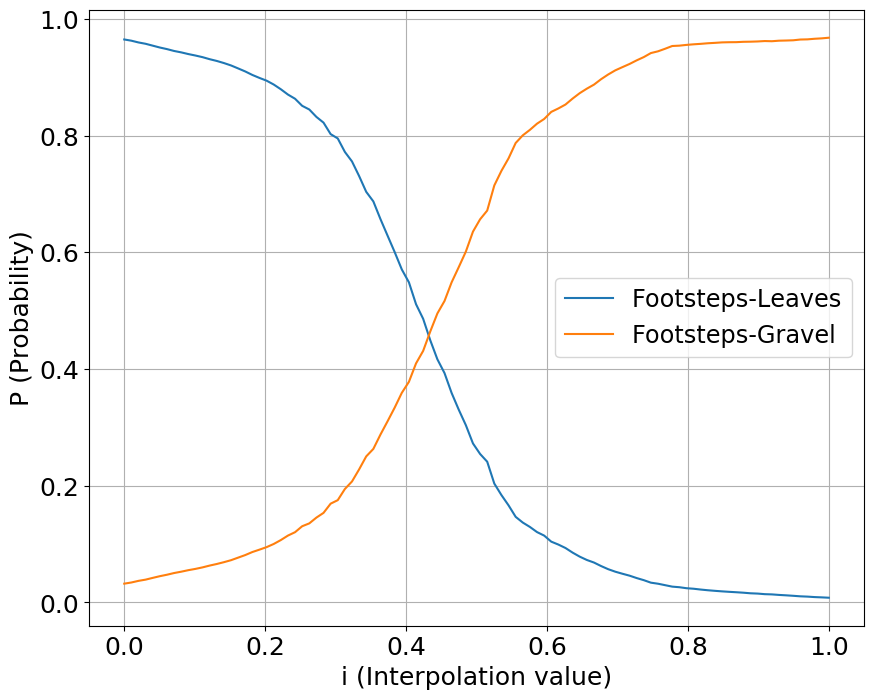}
    \end{subfigure}
    \caption{\textit{Interpolation for footsteps}}
  \end{subfigure}
  
  \vspace{0.3cm}
  
  % Subfigure for gunshots
  \begin{subfigure}[t]{\linewidth}
    \centering
    \begin{subfigure}[t]{0.22\linewidth}
      \includegraphics[width=\linewidth]{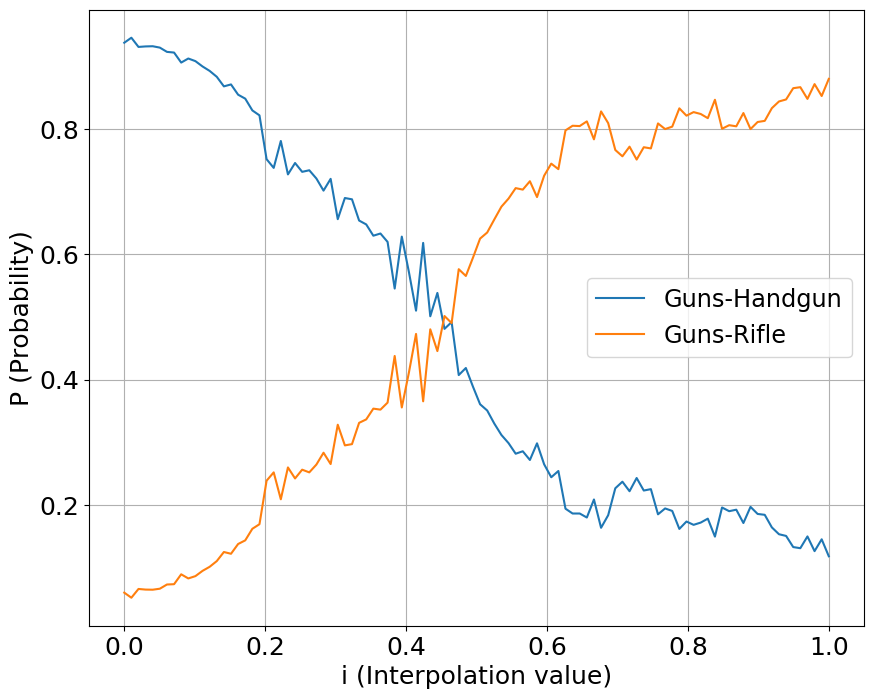}
    \end{subfigure}
    % \hspace{0.2cm} % Space between the images
    % \begin{subfigure}[t]{0.18\linewidth}
    %   \includegraphics[width=\linewidth]{interpolation/guns/guns-class-1-2-45-50.png}
    % \end{subfigure}
    \hspace{0.2cm} % Space between the images
    \begin{subfigure}[t]{0.22\linewidth}
      \includegraphics[width=\linewidth]{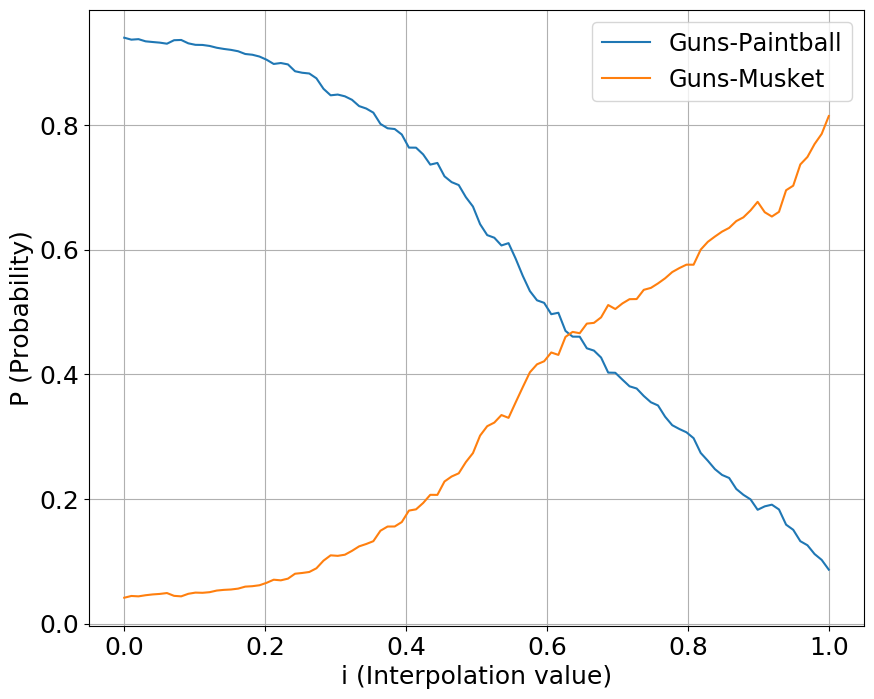}
    \end{subfigure}
    \hspace{0.2cm} % Space between the images
    \begin{subfigure}[t]{0.22\linewidth}
      \includegraphics[width=\linewidth]{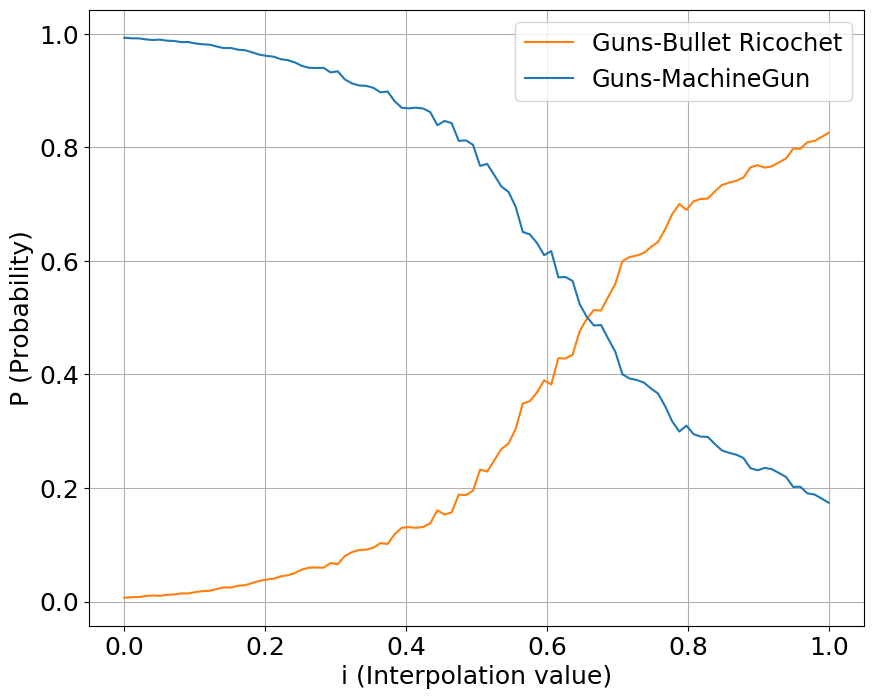}
    \end{subfigure}
    \hspace{0.2cm} % Space between the images
    \begin{subfigure}[t]{0.22\linewidth}
      \includegraphics[width=\linewidth]{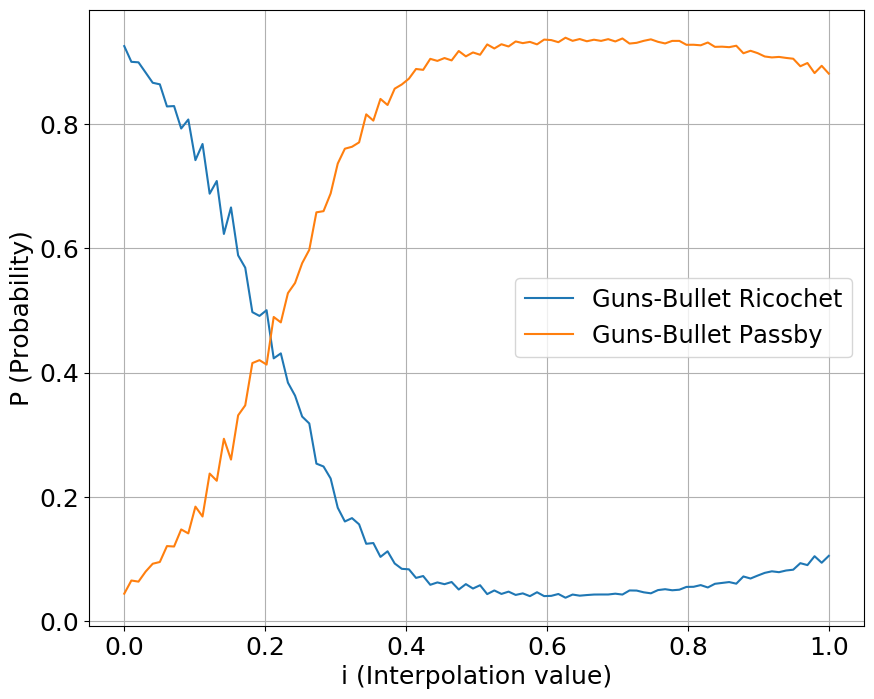}
    \end{subfigure}
    \caption{\textit{Interpolation for gunshots}}
  \end{subfigure}
  
  \vspace{0.3cm}
  
  % Subfigure for hits
  \begin{subfigure}[t]{\linewidth}
    \centering
    \begin{subfigure}[t]{0.22\linewidth}
      \includegraphics[width=\linewidth]{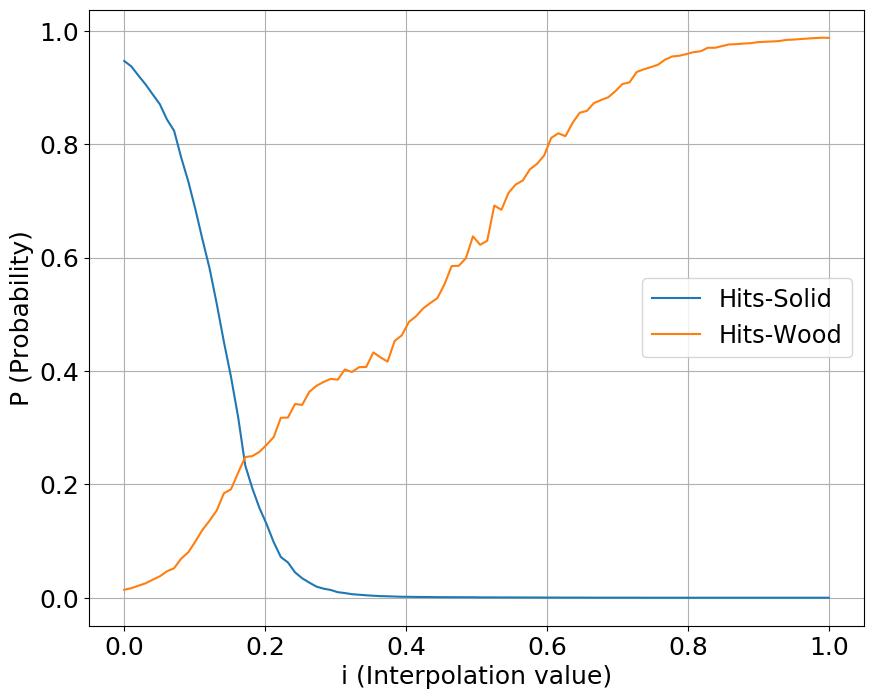}
    \end{subfigure}
    \hspace{0.2cm} % Space between the images
    % \begin{subfigure}[t]{0.18\linewidth}
    %   \includegraphics[width=\linewidth]{interpolation/hits/footstep-ht-class-1-6-12-15.png}
    % \end{subfigure}
    % \hspace{0.2cm} % Space between the images
    \begin{subfigure}[t]{0.22\linewidth}
      \includegraphics[width=\linewidth]{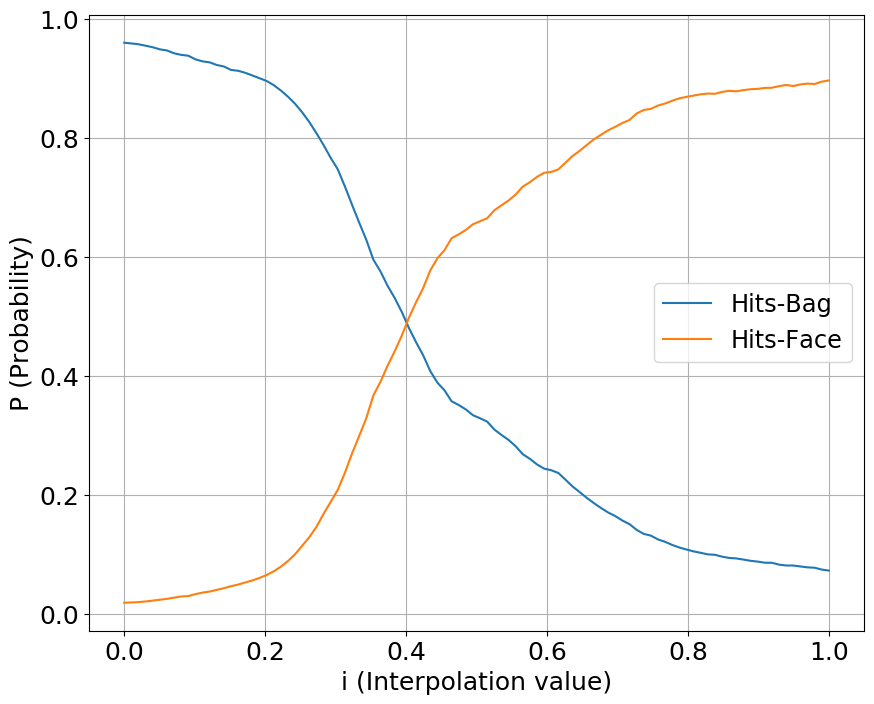}
    \end{subfigure}
    \hspace{0.2cm} % Space between the images
    \begin{subfigure}[t]{0.22\linewidth}
      \includegraphics[width=\linewidth]{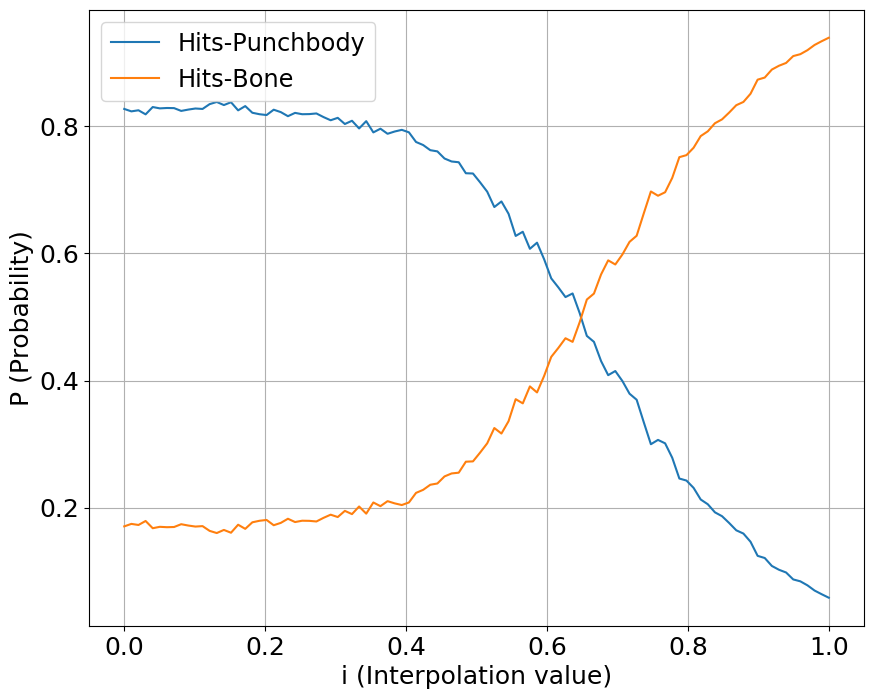}
    \end{subfigure}
    \hspace{0.2cm} % Space between the images
    \begin{subfigure}[t]{0.22\linewidth}
      \includegraphics[width=\linewidth]{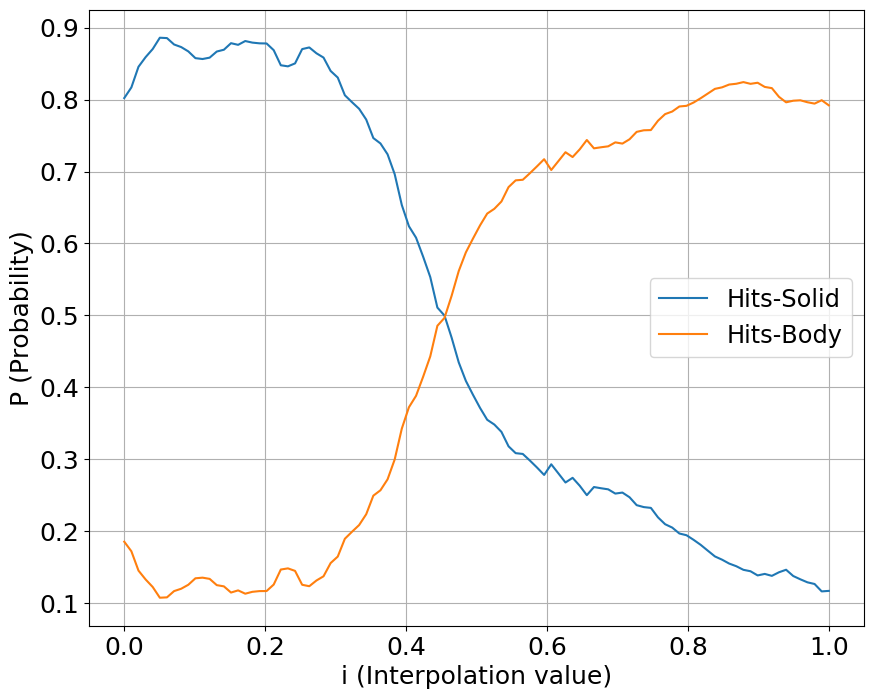}
    \end{subfigure}
    \caption{\textit{Interpolation for hits}}
  \end{subfigure}
  
  \caption{\textit{Effect of interpolating points in the conditioning space. We randomly selected four combinations of interpolation between two classes in each sound category. The horizontal axis represents the interpolation points between two targets, with one increasing from 0 to 1 and the other from 1 to 0. The vertical axis represents the probability for that class returned by a trained classifier.}}
  \label{fig: interpolation}
\end{figure*}

% \begin{figure}[h]
%   \caption{\textit{Effect of interpolating points in the conditioning space.}}
%   \centering
%   \centerline{\includegraphics[width=\linewidth]{Interpolation.png}}
% %  \vspace{1.5cm}
%   \label{fig: interpolation}
% \end{figure}

Once the model is fully trained, we can abandon both the encoder and the discriminator. Instead of extracting information from a sound's Mel spectrogram, we can define an arbitrary conditioning vector $C_s$. Entries in the $C_s$ vector close to unity indicate strong correspondence to a particular class within the sound type and values close to zero indicate no correspondence. In order to explore the influence of the conditioning vector,  we interpolate 100 points between two target classes within a sound type. In other words, the values for one class increase from zero to one, while the values for the other class decrease from one to zero. In order to evaluate the ``class'' of the sound we use a separate pre-trained classifier that is unrelated to the sound generation (described in the next section). In Figure~\ref{fig: interpolation}, we show four example pairs for each of the three sound categories. Each sound category was trained using a separate model with its own dataset. We randomly picked two classes within each category (eg, Footstep: concrete and gravel). Considering Figure~\ref{fig: interpolation}, we generally see smooth transitions from one class to another. Nonetheless, the sharpness of the transition seems to vary as does the completeness of the transition.

\subsection{Proposed metric for control effectiveness}

To measure the effectiveness of our conditioning method, we incorporate a pre-trained classifier that indicates the probability for each class of sound. This essentially links the conditioning information with the generated output. To this end, we use PANNs~\cite{PANNs}, a large-scale audio classification model trained on vast amounts of sounds~\cite{Audioset} with 14 CNN layers. It was reported to achieve great acoustic event detection and audio tagging performance agnostic to sound types. Based on the pre-trained PANNs model, we fine-tuned and trained it on our dataset until it converged and reached an evaluation score of 79.5\%.
\\
\\
We propose two evaluation metrics to evaluate our conditioning method: Maximum separation distance (MSD) and range of effective control (REC). The MSD is designed to quantify the breadth of the model's conditioning space. It does so by measuring the maximal extent to which the model's output can be varied by adjusting the conditioning vector to the limits indicating only one of the two target classes. In other words, ideally the conditioning vector can shift the generated sound completely from one class to the other. Practically, it can be incomplete. We measure the MSD score as the average of the total range of the softmax probability score for each of the two target classes:

\begin{equation}
    \text{MSD} = (P(j)_{max} - P(j)_{min} + P(k)_{max} - P(k)_{min}) / 2, 
\end{equation}
where $j, k$ are two interpolation variables with a range varying from zero to one.
\\
% \begin{equation}    
%     \text{MSD} = \frac{1}{\binom{n}{2}} \sum_{j=1}^{n-1} \sum_{k=j+1}^{n} \frac{P_j + P_k}{2}
% \end{equation}

The range of effective control (REC) is a metric that focuses on the practical operational range of the conditioning vector to change the model's output. It is calculated by determining the interval within which changes to the conditioning vector lead to significant changes in the output. More specifically, in Figure~\ref{fig: interpolation}, we see the curves can saturate as the interpolation values are varied towards the ends of the interpolation range. This is to say, the slope of the curves can become flat towards the ends.  Thus, we can define a threshold value for the absolute value or magnitude of the slope of the curve. If the magnitude of the slope is too small, less than some threshold, $\theta$, we say the response has saturated. We define the REC score as shown in equation~\ref{eq: REC}, $i\in [0,1]$.
\begin{equation}
\label{eq: REC}
    \text{REC} =  i_{\text{end}} - i_{\text{start}}
\end{equation}
where:
\begin{itemize}
    \item $i_{\text{start}}$ is the first interpolation value where $\left| \frac{\Delta P(i)}{\Delta i} \right| > \theta$,
    \item $i_{\text{end}}$ is the last interpolation value within the specified range where $\left| \frac{\Delta P(i)}{\Delta i} \right| > \theta$.
\end{itemize}
The REC score measures the segment of the output response curve for which the rate of change in the class probability is greater than the threshold value, $\theta$. We compute the REC score for all pairs of classes under consideration, to explore how effectively the interpolation between classes controls the timbre attributes of the sounds. 

\subsection{Evaluation setup}

In order to evaluate the synthesis performance of our model, we need to compare sounds generated by the model with real sounds. Therefore, we use our test dataset (refer to Section~\ref{sec: dataset}) to provide reference sounds. We extract the amplitude envelopes of the reference sounds and use a conditioning vector obtained by sampling a Gaussian model with a one-hot ground truth class label vector corresponding to the class of the reference sound as the mean for the Gaussian model.  We provide the amplitude envelope and conditioning vector as inputs to our ICGAN model and thus obtain one synthesized sound for each reference sound.  
\\
\\
We compare our ICGAN model with LTS (Latent timbre synthesis)~\cite{LTS}, which enables smooth transition from one sound to another by interpolating the VAE latent space. We denote ICGAN as the test in which we interpolated only conditioning vectors. Since LTS can only interpolate the latent space from one target sound to another and it does not take in any amplitude information as conditioning, for a fair comparison, we interpolated the amplitudes using ICGAN between two target sounds in the same way as LTS and named it ICGAN-w (with amplitude interpolation). For ablation study, we have also retrained our model by completely removing the amplitude envelope from the conditioning and named this ICGAN-n (no amplitude information). Finally, We added a comparison with the classical conditioning method as ~\cite{neuralfootstep}, namely, by inputting one-hot representations of the discrete labels into the generator and discriminator networks. We denote this model as conditional GAN (CGAN). Please note that we also interpolated values in the conditioning space of CGAN even though during training it was only accepting one-hot vectors. The reason for this is to test whether the model is able to extrapolate smooth transitions between discrete values. All results are shown in Table~\ref{table: Results}. We evaluate the quality of the synthesis by comparing the similarity between the synthesized sounds and the reference sounds. We select a range of evaluation metrics including Frechet Audio Distance (FAD)~\cite{FAD}, Frechet Inception Distance (FID)~\cite{FID}, and Log-Spectral Distance (LSD). Both FAD and FID are common metrics for evaluating the quality of generated data (audio and images) in generative models. They rely on pre-trained classifiers to extract features from both real and generated samples. The distance scores are then reported by computing the mean and covariance of these feature sets. This statistical approach captures not just the sound quality, but also the diversity within the datasets. For FAD, we use the VGGish~\cite{VGGish} classifier with a sample rate of 16kHz, and for FID, we employ the InceptionV3~\cite{InceptionV3} classifier. Additionally, we compute the pairwise log spectral distance between the generated Mel-spectrogram and the reference Mel-spectrogram. We report the final LSD as the average score from each category.

\section{Results}
\label{sec: Results}

\begin{table}[ht]
\centering
\resizebox{\columnwidth}{!}{%
\begin{tabular}{@{}l|lccccc@{}}
\toprule
\hline
& & \multicolumn{2}{c}{Control} & \multicolumn{3}{c}{Audio Quality} \\
Dataset & Model & MSD$\uparrow$ & REC$\uparrow$ & FAD$\downarrow$ & FID$\downarrow$ & LSD$\downarrow$\\ 
\midrule
\hline
\multirow{5}{*}{Footsteps} & ICGAN & 0.683 & 0.392 & \textbf{2.85} & \textbf{64.61} & 0.119\\
                           & ICGAN-w & 0.807 & \textbf{0.696} & / & / & /\\
                           & ICGAN-n & 0.744 & 0.610 & 16.66 & 163.4 & 0.118\\
                           & CGAN & \textbf{0.851} & 0.022 & 4.90 & 102.5 & \textbf{0.086}\\
                           & LTS & 0.759 & 0.667 & 8.86 & 145.6 & 0.125\\
\hline
\multirow{5}{*}{Guns}      & ICGAN & 0.637 & 0.579 & \textbf{2.64} & \textbf{60.44} & 0.235\\
                           & ICGAN-w & 0.733 & \textbf{0.721} & / & / & /\\
                           & ICGAN-n & 0.722 & 0.629 & 7.09 & 177.4 & 0.197\\
                           & CGAN & \textbf{0.841} & 0.018 & 2.88 & 76.43 & \textbf{0.187}\\
                           & LTS & 0.637 & 0.711 & 4.97 & 126.9 & 0.289\\
\hline
\multirow{5}{*}{Hits}      & ICGAN & 0.647 & 0.863 & \textbf{1.27} & \textbf{60.58} & 0.136\\
                           & ICGAN-w & 0.795 & 0.903 & / & / & /\\
                           & ICGAN-n & 0.773 & 0.807 & 5.19 & 110.6 & 0.146\\
                           & CGAN & \textbf{0.911} & 0.011 & 2.43 & 95.57 & \textbf{0.124}\\
                           & LTS & 0.677 & \textbf{0.936} & 2.93 & 97.87 & 0.176\\
\hline
\multirow{5}{*}{DCASE}     & ICGAN & 0.509 & 0.210 & \textbf{6.956} & \textbf{79.60} & 0.272\\
                           & ICGAN-w & 0.727 & \textbf{0.569} & / & / & /\\
                           & ICGAN-n & 0.641 & 0.513 & 16.49 & 185.5 & 0.292\\
                           & CGAN & \textbf{0.818} & 0.026 & 10.21 & 108.6 & \textbf{0.246}\\
                           & LTS & 0.693 & 0.512 & 7.67 & 131.4 & 0.321\\
\bottomrule
\hline
\end{tabular}%
}
\caption{Comparison of model synthesis performance and control affordances. ICGAN is the proposed model as illustrated in \ref{fig: model}. ICGAN refers to our proposed method by interpolating only conditioning space. ICGAN-w is the same ICGAN model but we also interpolated the extracted amplitudes from two target sounds. Therefore the audio quality is the same as ICGAN. ICGAN-n is a re-trained model which completely removes amplitude information. CGAN denotes the conditional GAN using concatenation. LTS refers to 'Latent Timbre Synthesis'~\cite{LTS} which does latent space interpolation between two target sounds. Up-arrows indicate the higher the score the better, and vice versa.}
\label{table: Results}
\end{table}

\subsection{Conditioning performance}

% \begin{figure}[ht]
%   \centering
%   \centerline{\includegraphics[width=\linewidth]{REC.png}}
%   \caption{Range of effective control}
%   \label{fig: conditional}
% \end{figure}

% \begin{table}[ht]
% \centering
% \resizebox{\columnwidth}{!}{%
% \begin{tabular}{c|cc|cc|cc}
%    \hline
%    \multirow{2}{*}{Models}  & \multicolumn{2}{c|}{Footsteps} & \multicolumn{2}{c|}{Guns} & \multicolumn{2}{c}{Hits} \\
%        & MSD\uparrow & REC\uparrow & MSD\uparrow & REC\uparrow & MSD\uparrow & REC\uparrow \\
%    \hline
%    ICGAN & 0.683 & 0.667 & 0.637 & 0.711 & 0.647 & 0.807\\
%    ICGAN-WL & & & & & & \\
%    LTS & 0.392 & 0.430 & 0.579 & 0.663  & 0.863 & 0.859\\
%    \hline
% \end{tabular}%
% }
% \caption{Evaluation of Maximum Separation Distance and Range of Effective Encoding}
% \label{tab:objective}
% \end{table}

Since we have several classes of sounds for each sound type or category, there are several possible pairs of classes with which to interpolate between. We refer to these pairs of classes as interpolation pairs. For the purposes of showing results, we average the results across a sample of interpolation pairs.  Average results obtained for the MSD and REC are shown in Table~\ref{table: Results}. When computing the REC, we use a threshold value given by $\theta = 1e^{-2}$. For the MSD score, we found that CGAN achieves the highest score. This makes sense because it was hard conditioned on discrete values between two ends, namely 0 and 1. The classification could successfully tell the class on the two ends. On the contrary, REC reflects that CGAN gets much lower scores because the transition between two ends is rather discrete and drastic. Our ICGAN-w, which we interpolated the amplitudes together with the conditioning space, achieves the highest REC overall, and performs second best in MSD. This shows the potential of our method, as it no longer requires interpolating in the uninterpretable latent space to shift from one sound to another. Although by interpolating the conditioning space only did not yield the best performance, we nonetheless found it useful as it provides a straightforward and interpretable control over the sound categories, which could not be easily achieved using other models relying on extensive labels. 
\\
\\
Additionally, the results for the REC score indicate that the effective range of conditioning varies across different sound types. We found the Hits category obtains the highest REC overall, whereas DCASE dataset attains the lowest. We hypothesize that this can be attributed to the degree of variance within different sound classes. To test this, we extract and average the MFCC features for each sub-category, then compute and normalize the pairwise Euclidean distances between categories, and finally get the averaged similarity scores for each dataset. We found that our Hits dataset did have higher variances among different sub-categories, with the normalized similarity score 0.568, which is higher than Guns 0.451, Footsteps 0.316 and DCASE 0.342. The high degree of variations among different sub-categories could enable the encoder classifier to efficiently learn the intricate differences and thereby output meaningful and smooth values, which likely results in the smooth transitions different classes. However, how exactly dataset variations could contribute to different control affordances needs further investigation.

% We noticed   The MSD score is limited by several factors such as the accuracy of the pre-trained audio classifier and the similarity of the sounds in each class. The MSD scores indicate reasonable performance in that it is agnostic to the sound types and reaches an average greater than $0.63$. In the future work, we could perform a more accurate evaluation with a dedicated classifier to distinguish the sound timbres agnostic to amplitude envelope. 

\subsection{Synthesis performance}
\label{sec: inference}

% \caption{\textit{Examples showing how interpolating conditioning space contributes to spectral variations. For more examples and to listen to the sounds, please refer to our accompanying website~\footnote{\url{https://reinliu.github.io/ICGAN-Paper/}}. }}

In Table~\ref{table: Results}, we found that our model excels in statistical similarity metrics reflected in FAD and FID scores, indicating a high degree of performance in the quality and diversity of the generated data. However, from the log spectral distances which compare the pairwise signals directly, we realized the explicit conditioning method achieves higher accuracy in the spectrogram reconstruction. This is because our implicit conditioning method introduces uncertainty through sampling, resulting in lower accuracy for reconstructing signals with definitive labels. On the contrary, this also brings higher sample diversities reflected in the statistical measures. We also noticed an increase in dissimilarity between the target and generated data trained by the DCASE Foley dataset. It becomes more difficult for both models to reconstruct the signals when the conditioned data vary more. We have also observed that after removing the extracted amplitudes, the audio quality was reduced significantly. We believe the amplitude information provides an important guidance for the generator for synthesis.

\begin{figure}[ht]
    \centering
    \begin{subfigure}[t]{\linewidth}
      \includegraphics[width=\linewidth]{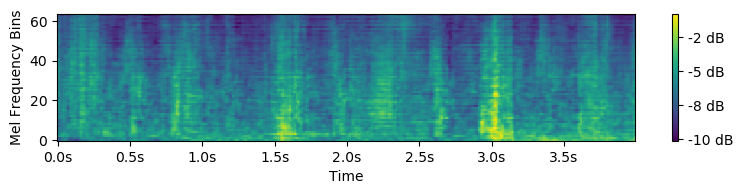}
      \caption{Footstep-concrete}
      \label{fig: Footstep-concrete}
    \end{subfigure}
    \begin{subfigure}[t]{\linewidth}
      \includegraphics[width=\linewidth]{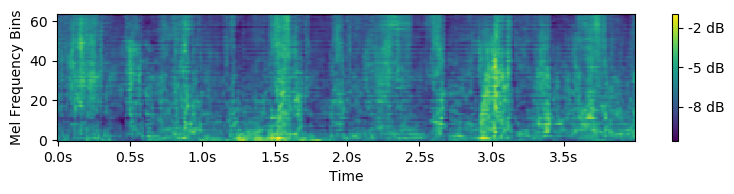}
      \caption{Footstep-leaves}
      \label{fig: Footstep-leaves}
    \end{subfigure}
    \begin{subfigure}[t]{\linewidth}
      \includegraphics[width=\linewidth]{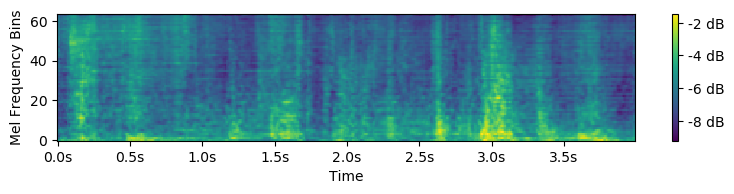}
      \caption{Footstep-gravel}
      \label{fig: Footstep-gravel}
    \end{subfigure}
    \caption{\textit{Generated in-class sounds. The models are trained in the same footsteps category with different variations denoted by its property.}}
    \label{fig: In-domain}
\end{figure}

In Figure~\ref{fig: In-domain}, we show the synthesized Mel spectrograms corresponding to different classes within the Foodsteps sound category: concrete, leaves, and gravel. Note that all three sounds were generated using the same amplitude envelope, and the spectrograms exhibited similar temporal changes. The amplitude conditioning helps the model maintain the overall contours of the spectrogram while offering controllability varying timbral characteristics. In addition to conditioning on in-class sounds, our model could also be trained to synthesize out-of-domain sounds. We show the results of cross-domain sounds trained with DCASE dataset in our accompaniment website~\footnote{\url{https://reinliu.github.io/ICGAN-Paper/}}. 
% In other words, the conditioning vector now represents a much greater variation in spectral and timbral quality. For example, we can consider a conditioning vector that spans footstep sounds~\ref{fig: footstep}, gunshot sounds~\ref{fig: Gunshot}, and dogbark sounds~\ref{fig: Dogbark}. While the spectral characteristics vary among the three categories of sounds, the sound generator often fails to output high-fidelity sounds. This can likely be attributed to the fact that the amplitude envelope is over constraining the model. Therefore, in future work we hope to explore larger datasets with more variances in the temporal and spectral information.

% \begin{figure}
%     \centering
%     \begin{subfigure}[t]{\linewidth}
%       \includegraphics[width=\linewidth]{images/dev-footstep-0.png}
%       \caption{Footstep}
%       \label{fig: footstep}
%     \end{subfigure}
    
%     \begin{subfigure}[t]{\linewidth}
%       \includegraphics[width=\linewidth]{images/dev-gun-0.png}
%       \caption{Gunshot}
%       \label{fig: Gunshot}
%     \end{subfigure}
    
%     \begin{subfigure}[t]{\linewidth}
%       \includegraphics[width=\linewidth]{images/dev-dog-0.png}
%       \caption{Dogbark}
%       \label{fig: Dogbark}
%     \end{subfigure}
%     \label{fig: generated-group}
%     \caption{\textit{Cross-domain sound spectrograms. We trained the single model conditioned on different classes using the DCASE~\cite{DCASE} dataset.}}
%   \label{fig: cross-domain}
% \end{figure}

\section{Discussion and Conclusion}
\label{sec: Discussion}

In this research, we proposed a novel conditioning method that translates discrete labels into a continuous probabilistic labelling space. We have applied this implicit conditioning method to the domain of neural audio synthesis and explored model performance in terms of the effectiveness of the conditioning vector in controlling the sound class and the audio quality of the generated sounds. We primarily focused on in-domain variations in sound effects, but briefly consider cross-domain sound effects. Our approach shows promise when compared with conventional conditioning methods, but requires more development. We show that the conditioning vector can effectively interpolate between sound classes and enable a smooth transition in the timbre space without relying on excessive labels. We also observed that the control effectiveness seems to relate to the variations of a dataset, but we will leave this question to future investigation. We hope our proposed conditioning approach sparks more creativity in the sound design and synthesis field and allows for interesting future explorations in the continuous conditioning space, such as interpolation involving more than two dimensions and also mixing the features of various target sounds. Future work could include incorporating a training mechanism that encourages a more consistent and controllable conditioning space.

%\newpage
\nocite{*}
\bibliographystyle{IEEEbib}
\bibliography{ICGAN}

\begin{thebibliography}{10}

\bibitem{wavenet}
A.~van~den Oord, S.~Dieleman, H.~Zen, K.~Simonyan, O.~Vinyals, A.~Graves, N.~Kalchbrenner, A.~W. Senior, and K.~Kavukcuoglu,
\newblock ``Wavenet: A generative model for raw audio,''
\newblock in {\em The 9th ISCA Speech Synthesis Workshop}, Sunnyvale, CA, USA, September 2016, p. 125, ISCA.

\bibitem{MelGAN}
Kundan Kumar, Rithesh Kumar, Thibault de~Boissi{\`e}re, Lucas Gestin, Wei~Zhen Teoh, Jose M.~R. Sotelo, Alexandre de~Br{\'e}bisson, Yoshua Bengio, and Aaron~C. Courville,
\newblock ``Melgan: Generative adversarial networks for conditional waveform synthesis,''
\newblock in {\em Neural Information Processing Systems}, 2019.

\bibitem{RAVE}
Antoine Caillon and Philippe Esling,
\newblock ``Rave: A variational autoencoder for fast and high-quality neural audio synthesis,'' 2021.

\bibitem{GANsynth}
Jesse Engel, Kumar~Krishna Agrawal, Shuo Chen, Ishaan Gulrajani, Chris Donahue, and Adam Roberts,
\newblock ``Gansynth: Adversarial neural audio synthesis,''
\newblock in {\em International Conference on Learning Representations}, 2019, vol. abs/1902.08710.

\bibitem{conwavegan}
Adri{\'a}n Barahona-R{\i}os and Sandra Pauletto,
\newblock ``Synthesising knocking sound effects using conditional wavegan,''
\newblock in {\em 17th Sound and Music Computing Conference, Online}, 2020.

\bibitem{neuralfootstep}
Marco Comunit\`{a}, Huy Phan, and Joshua~D. Reiss,
\newblock ``Neural synthesis of footsteps sound effects with generative adversarial networks,''
\newblock in {\em Audio Engineering Society Convention 152}, Online and In-Person, May 2022, 2022, Audio Engineering Society,
\newblock Convention Paper 10583.

\bibitem{liu2023conditional}
Yunyi Liu and Craig Jin,
\newblock ``Conditional sound effects generation with regularized wgan,''
\newblock in {\em 20th Sound and Music Computing Conference, Stockholm}, 2023.

\bibitem{DarkGAN}
Javier Nistal, Stefan Lattner, and Ga{\"e}l Richard,
\newblock ``Darkgan: Exploiting knowledge distillation for comprehensible audio synthesis with gans,''
\newblock in {\em International Society for Music Information Retrieval Conference}, 2021.

\bibitem{PANNs}
Qiuqiang Kong, Yin Cao, Turab Iqbal, Yuxuan Wang, Wenwu Wang, and Mark~D. Plumbley,
\newblock ``Panns: Large-scale pretrained audio neural networks for audio pattern recognition,''
\newblock {\em IEEE/ACM Transactions on Audio, Speech, and Language Processing}, vol. 28, pp. 2880--2894, 2019.

\bibitem{10.5555/3157382.3157633}
A\"{a}ron van~den Oord, Nal Kalchbrenner, Oriol Vinyals, Lasse Espeholt, Alex Graves, and Koray Kavukcuoglu,
\newblock ``Conditional image generation with pixelcnn decoders,''
\newblock in {\em Proceedings of the 30th International Conference on Neural Information Processing Systems}, Red Hook, NY, USA, 2016, NIPS'16, p. 4797–4805, Curran Associates Inc.

\bibitem{constraints}
Zhiting Hu, Zichao Yang, Ruslan Salakhutdinov, Xiaodan Liang, Lianhui Qin, Haoye Dong, and Eric~P. Xing,
\newblock ``Deep generative models with learnable knowledge constraints,''
\newblock in {\em Proceedings of the 32nd International Conference on Neural Information Processing Systems}, Red Hook, NY, USA, 2018, NIPS'18, p. 10522–10533, Curran Associates Inc.

\bibitem{Vrtes2018FlexibleAA}
Eszter V{\'e}rtes and Maneesh Sahani,
\newblock ``Flexible and accurate inference and learning for deep generative models,''
\newblock in {\em Neural Information Processing Systems}, 2018.

\bibitem{hierarchical}
Yunchuan Guan, Yu~Liu, Ke~Zhou, and Junyuan Huang,
\newblock ``Hierarchical meta-learning with hyper-tasks for few-shot learning,''
\newblock in {\em Proceedings of the 32nd ACM International Conference on Information and Knowledge Management}, New York, NY, USA, 2023, CIKM '23, p. 587–596, Association for Computing Machinery.

\bibitem{AudioLDM}
Haohe Liu, Zehua Chen, Yiitan Yuan, Xinhao Mei, Xubo Liu, Danilo~P. Mandic, Wenwu Wang, and MarkD~. Plumbley,
\newblock ``Audioldm: Text-to-audio generation with latent diffusion models,''
\newblock in {\em International Conference on Machine Learning}, 2023.

\bibitem{conaug}
Heewoo Jun, Rewon Child, Mark Chen, John Schulman, Aditya Ramesh, Alec Radford, and Ilya Sutskever,
\newblock ``Distribution augmentation for generative modeling,''
\newblock in {\em Proceedings of the 37th International Conference on Machine Learning}. 2020, ICML'20, JMLR.org.

\bibitem{CCGAN}
Xin Ding, Yongwei Wang, Zuheng Xu, William~J. Welch, and Z.~Jane Wang,
\newblock ``Ccgan: Continuous conditional generative adversarial networks for image generation,''
\newblock {\em CoRR}, vol. abs/2011.07466, 2020.

\bibitem{PcDGAN}
Amin Heyrani~Nobari, Wei Chen, and Faez Ahmed,
\newblock ``Pcdgan: A continuous conditional diverse generative adversarial network for inverse design,''
\newblock in {\em Proceedings of the 27th ACM SIGKDD Conference on Knowledge Discovery \& Data Mining}, New York, NY, USA, 2021, KDD '21, p. 606–616, Association for Computing Machinery.

\bibitem{Labelsmoothing}
Chang-Bin Zhang, Peng-Tao Jiang, Qibin Hou, Yunchao Wei, Qi~Han, Zhen Li, and Ming-Ming Cheng,
\newblock ``Delving deep into label smoothing,''
\newblock vol. 30, pp. 5984–5996, jan 2021.

\bibitem{WGAN}
Mart{\'i}n Arjovsky, Soumith Chintala, and L{\'e}on Bottou,
\newblock ``Wasserstein generative adversarial networks,''
\newblock in {\em International Conference on Machine Learning}, 2017.

\bibitem{VAE}
Diederik~P. Kingma and Max Welling,
\newblock ``Auto-encoding variational bayes,''
\newblock {\em International Conference on Learning Representations}, vol. abs/1312.6114, 2013.

\bibitem{FILM}
Ethan Perez, Florian Strub, Harm de~Vries, Vincent Dumoulin, and Aaron Courville,
\newblock ``Film: visual reasoning with a general conditioning layer,''
\newblock in {\em Proceedings of the Thirty-Second AAAI Conference on Artificial Intelligence and Thirtieth Innovative Applications of Artificial Intelligence Conference and Eighth AAAI Symposium on Educational Advances in Artificial Intelligence}. 2018, AAAI'18/IAAI'18/EAAI'18, AAAI Press.

\bibitem{VQVAE}
Xubo Liu, Turab Iqbal, Jinzheng Zhao, Qiushi Huang, Mark~D. Plumbley, and Wenwu Wang,
\newblock ``Conditional sound generation using neural discrete time-frequency representation learning,''
\newblock {\em 2021 IEEE 31st International Workshop on Machine Learning for Signal Processing (MLSP)}, pp. 1--6, 2021.

\bibitem{Specsingan}
Adrian Barahona~Rios and Tom Collins,
\newblock ``Specsingan: Sound effect variation synthesis using single-image gans,''
\newblock in {\em Proceedings of the Sound and Music Computing Conference}, 2022.

\bibitem{Hifigan}
Jungil Kong, Jaehyeon Kim, and Jaekyoung Bae,
\newblock ``Hifi-gan: generative adversarial networks for efficient and high fidelity speech synthesis,''
\newblock in {\em Proceedings of the 34th International Conference on Neural Information Processing Systems}, Red Hook, NY, USA, 2020, NIPS'20, Curran Associates Inc.

\bibitem{STGAN}
Yunyi {Liu} and Craig {Jin},
\newblock ``{Impact on quality and diversity from integrating a reconstruction loss into neural audio synthesis},''
\newblock {\em Acoustical Society of America Journal}, vol. 154, pp. A99--A99, Oct. 2023.

\bibitem{LENET}
Y.~Lecun, L.~Bottou, Y.~Bengio, and P.~Haffner,
\newblock ``Gradient-based learning applied to document recognition,''
\newblock {\em Proceedings of the IEEE}, vol. 86, no. 11, pp. 2278--2324, 1998.

\bibitem{Audioset}
Jort~F. Gemmeke, Daniel P.~W. Ellis, Dylan Freedman, Aren Jansen, Wade Lawrence, R.~Channing Moore, Manoj Plakal, and Marvin Ritter,
\newblock ``Audio set: An ontology and human-labeled dataset for audio events,''
\newblock in {\em Proc. IEEE ICASSP 2017}, New Orleans, LA, 2017.

\bibitem{WGAN-GP}
Ishaan Gulrajani, Faruk Ahmed, Martin Arjovsky, Vincent Dumoulin, and Aaron~C Courville,
\newblock ``Improved training of wasserstein gans,''
\newblock {\em Advances in neural information processing systems}, vol. 30, 2017.

\bibitem{BBC}
{Films for the Humanities \& Sciences},
\newblock {\em BBC Sound Effects Library},
\newblock Princeton, N.J., 1991,
\newblock Sound effects library.

\bibitem{ADAM}
Diederik~P. Kingma and Jimmy Ba,
\newblock ``Adam: {A} method for stochastic optimization,''
\newblock in {\em 3rd International Conference on Learning Representations, {ICLR} 2015, San Diego, CA, USA, May 7-9, 2015, Conference Track Proceedings}, Yoshua Bengio and Yann LeCun, Eds., 2015.

\bibitem{DCASE}
Keunwoo Choi, Jaekwon Im, Laurie Heller, Brian McFee, Keisuke Imoto, Yuki Okamoto, Mathieu Lagrange, and Shinosuke Takamichi,
\newblock ``Foley sound synthesis at the dcase 2023 challenge,''
\newblock {\em In arXiv e-prints: 2304.12521}, 2023.

\bibitem{LTS}
Kıvanç Tatar, Daniel Bisig, and Philippe Pasquier,
\newblock ``Latent timbre synthesis,''
\newblock {\em Neural Computing and Applications}, vol. 33, pp. 1--18, 01 2021.

\bibitem{FAD}
Kevin Kilgour, Mauricio Zuluaga, Dominik Roblek, and Matthew Sharifi,
\newblock ``Fr{\'e}chet audio distance: A reference-free metric for evaluating music enhancement algorithms,''
\newblock in {\em Interspeech}, 2019.

\bibitem{FID}
Martin Heusel, Hubert Ramsauer, Thomas Unterthiner, Bernhard Nessler, and Sepp Hochreiter,
\newblock ``Gans trained by a two time-scale update rule converge to a local nash equilibrium,''
\newblock in {\em Proceedings of the 31st International Conference on Neural Information Processing Systems}, Red Hook, NY, USA, 2017, NIPS'17, p. 6629–6640, Curran Associates Inc.

\bibitem{VGGish}
Shawn Hershey, Sourish Chaudhuri, Daniel P.~W. Ellis, Jort~F. Gemmeke, Aren Jansen, R.~Channing Moore, Manoj Plakal, Devin Platt, Rif~A. Saurous, Bryan Seybold, Malcolm Slaney, Ron~J. Weiss, and Kevin~W. Wilson,
\newblock ``Cnn architectures for large-scale audio classification,''
\newblock {\em 2017 IEEE International Conference on Acoustics, Speech and Signal Processing (ICASSP)}, pp. 131--135, 2016.

\bibitem{InceptionV3}
Christian Szegedy, Vincent Vanhoucke, Sergey Ioffe, Jonathon Shlens, and Zbigniew Wojna,
\newblock ``Rethinking the inception architecture for computer vision,''
\newblock {\em 2016 IEEE Conference on Computer Vision and Pattern Recognition (CVPR)}, pp. 2818--2826, 2015.

\bibitem{LDM}
Robin Rombach, A.~Blattmann, Dominik Lorenz, Patrick Esser, and Bj{\"o}rn Ommer,
\newblock ``High-resolution image synthesis with latent diffusion models,''
\newblock {\em 2022 IEEE/CVF Conference on Computer Vision and Pattern Recognition (CVPR)}, pp. 10674--10685, 2021.

\end{thebibliography}

% \section{Appendix}
% \label{sec: appendix}
% % This section shows the column margins for the text. \bigskip\newline

% \begin{figure}[ht]
%   \centering
%     \includegraphics[width=0.8\linewidth]{images/generator.png}
%     \caption{Generator Architecture}
%     \label{fig:Generator}
% \end{figure}

% \begin{figure}[ht]
%     \centering
%     \includegraphics[width=0.7
% \linewidth]{images/discriminator.png}
%     \caption{Discriminator Architecture}
%     \label{fig:discriminator}
% \end{figure}

\end{document}